\documentclass[reprint,superscriptaddress,amsmath,amssymb,aps,prb]{revtex4-1}

\usepackage{dcolumn}
\usepackage{graphicx}
\usepackage{hyperref}
\usepackage{color}
\usepackage[english]{babel}
\usepackage{bm}
\usepackage[T1]{fontenc}

\begin{document}

\begin{titlepage}

{\fontsize{26}{10} \textbf{\textcolor{black}{\flushleft Measurement and microscopic description of odd-even staggering of charge radii of exotic copper isotopes}}}\\

{
R.~P.~de Groote $^{1,2,*}$,
J.~Billowes $^{3}$,
C.~L.~Binnersley $^{3}$,
M.~L.~Bissell $^{3}$,
T.~E.~Cocolios $^{1}$,
T. Day Goodacre $^{4,5}$,
G.~J.~Farooq-Smith $^{1}$,
D.~V.~Fedorov $^{6}$,
K.~T.~Flanagan $^{3}$,
S.~Franchoo $^{7}$,
R.~F.~Garcia Ruiz $^{3,8,9}$,
W.~Gins $^{1,2}$,
J.D.~Holt $^{5}$,
\'{A}.~Koszor\'{u}s $^{1}$,
K.~M.~Lynch $^{9}$,
T. Miyagi $^{5}$,
W.~Nazarewicz $^{10}$,
G.~Neyens $^{1,9}$,
P.-G.~Reinhard $^{11}$,
S.~Rothe $^{3,4}$,
H.~H.~Stroke $^{12}$,
A.~R.~Vernon $^{1,3}$,
K.~D.~A.~Wendt $^{13}$,
S.~G.~Wilkins $^{3,4}$,
Z.~Y.~Xu $^{1}$ and 
X.~F.~Yang $^{1,14}$
}

{
\fontsize{8}{10}{
\selectfont
$^{1}$ KU Leuven, Instituut voor Kern- en Stralingsfysica, B-3001 Leuven, Belgium,
$^{2}$ Department of Physics, University of Jyv\"askyl\"a, PB 35(YFL) FIN-40351 Jyv\"askyl\"a, Finland,
$^{3}$ Photon Science Institute, Department of Physics and Astronomy, The University of Manchester, Manchester M13 9PL, UK,
$^{4}$ Engineering Department, CERN, CH-1211 Geneva 23, Switzerland,
$^{5}$ TRIUMF, 4004 Wesbrook Mall, Vancouver, BC V6T 2A3, Canada,
$^{6}$ Petersburg Nuclear Physics Institute, 188350 Gatchina, Russia,
$^{7}$ Institut de Physique Nucl\'{e}aire d'Orsay, F-91406 Orsay, France,
$^{8}$ Massachusetts Institute of Technology, Cambridge, MA 02139, USA,
$^{9}$ Experimental Physics Department, CERN, CH-1211 Geneva 23, Switzerland,
$^{10}$ Department of Physics and Astronomy and FRIB Laboratory, Michigan State University, East Lansing, Michigan 48824, USA,
$^{11}$ Institut f\"{u}r Theoretische Physik, Universit\"{a}t Erlangen, D-91054 Erlangen, Germany,
$^{12}$ Department of Physics, New York University, New York, New York 10003, USA,
$^{13}$ Institut f\"{u}r Physik, Johannes Gutenberg-Universit\"{a}t, D-55128 Mainz, Germany,
$^{14}$ School of Physics and State Key Laboratory of Nuclear Physics and Technology, Peking University, Beijing 100871, China
}
}
\vspace{0.1cm}
\end{titlepage}

\textbf{
The mesoscopic nature of the atomic nucleus gives rise to a wide array of macroscopic and microscopic phenomena. The size of the nucleus is a window into this duality: while the charge radii globally scale as $\bm{A^{1/3}}$, their evolution across isotopic chains reveals unanticipated structural phenomena~\cite{GarciaRuiz2016,Mil18a,marsh2018}. 
The most ubiquitous of these is perhaps the Odd-Even Staggering (OES)~\cite{Reehal1971}: isotopes with an odd number of neutrons are usually smaller in size than the trend of their even-neutron neighbours suggests. 
This OES effect varies with the number of protons and neutrons and poses a significant challenge for nuclear theory~\cite{reinhard2016,hammen2018,gorges2019}.
Here, we examine this problem with new measurements of the charge radii of short-lived copper isotopes up to the very exotic $^{78}$Cu ($\bm{Z=29, N=49}$), produced at only 20 ions/s, using the highly-sensitive Collinear Resonance Ionisation Spectroscopy (CRIS) method at ISOLDE-CERN. Due to the presence of a single proton outside of the closed $\bm{Z=28}$ shell, these measurements provide crucial insights into the single-particle proton structure and how this affects the charge radii. We observe an unexpected reduction in the OES for isotopes approaching the $\bm{N=50}$ shell gap. 
To describe the data, we applied models based on nuclear Density Functional Theory~\cite{Reinhard2017,Mil18a} (DFT) and ab-initio Valence-Space In-Medium Similarity Renormalization Group (VS-IMSRG) theory~\cite{Tsuk12SM,Stro19ARNPS}. Through these comparisons, we demonstrate a relation between the global behavior of charge radii and the saturation density of nuclear matter, and show that the local charge radii variations, which reflect the many-body polarization effects due to the odd neutron, naturally emerge from the VS-IMSRG calculations.
}


The properties of exotic nuclei, in particular of those close to (doubly) magic systems far from stability, have continually proven pivotal in deepening our understanding of nuclear forces and many-body dynamics. Until now, experimentally accessing charge radii within one or two nucleons of exotic doubly magic systems like $^{78}$Ni and $^{100}$Sn has proven prohibitively difficult. The most widely applied methods for measuring charge radii of short-lived radioactive isotopes rely on the fact that changes of the charge radius of the nucleus result in small changes in the energies of the atomic electrons, which is typically a 1:$10^6$ effect. These isotope shifts, while small, can be measured using high-resolution laser spectroscopy techniques, and be used to determine the changes in the mean-squared charge radius (see the Methods section for more details). Due to the presence of the unpaired proton, odd-$Z$ isotopes like the copper isotopes we study here, provide crucial insights into the single-particle proton structure and how this affects the charge radii. So far, the charge radii of copper isotopes could only be extracted up to $^{75}$Cu $(N=46)$~\cite{Bissell2016}. 

\begin{figure}[ht]
  \includegraphics[width = 1\columnwidth]{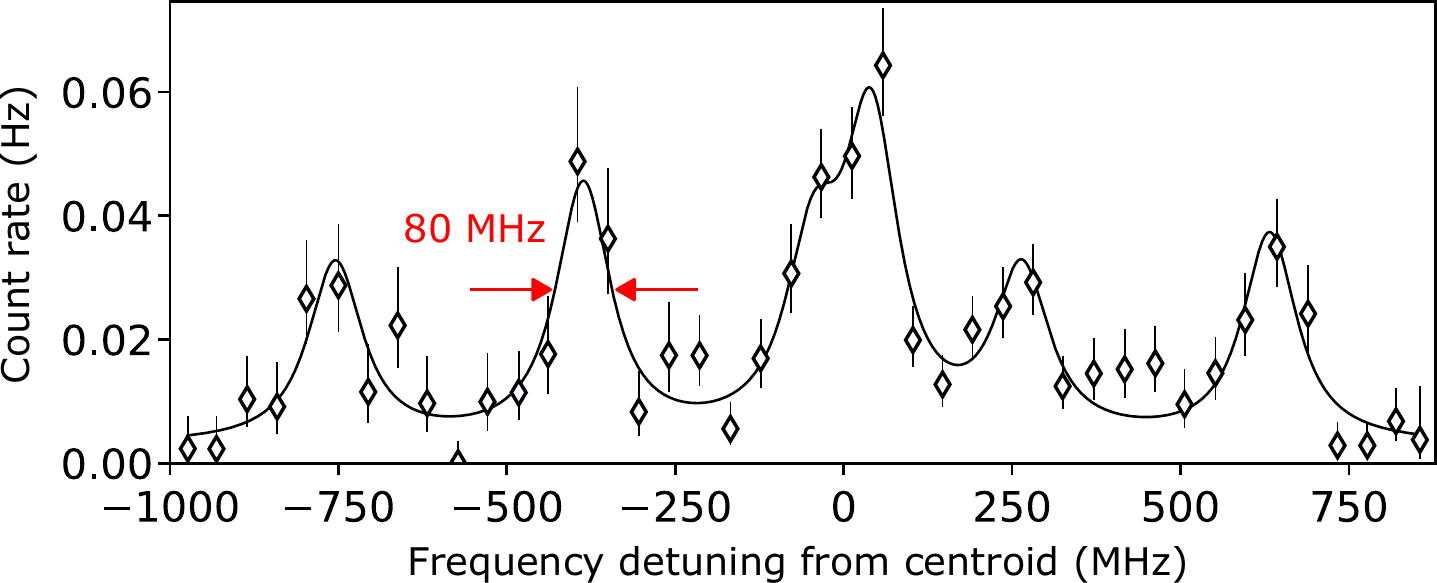}
  \includegraphics[width = 1\columnwidth]{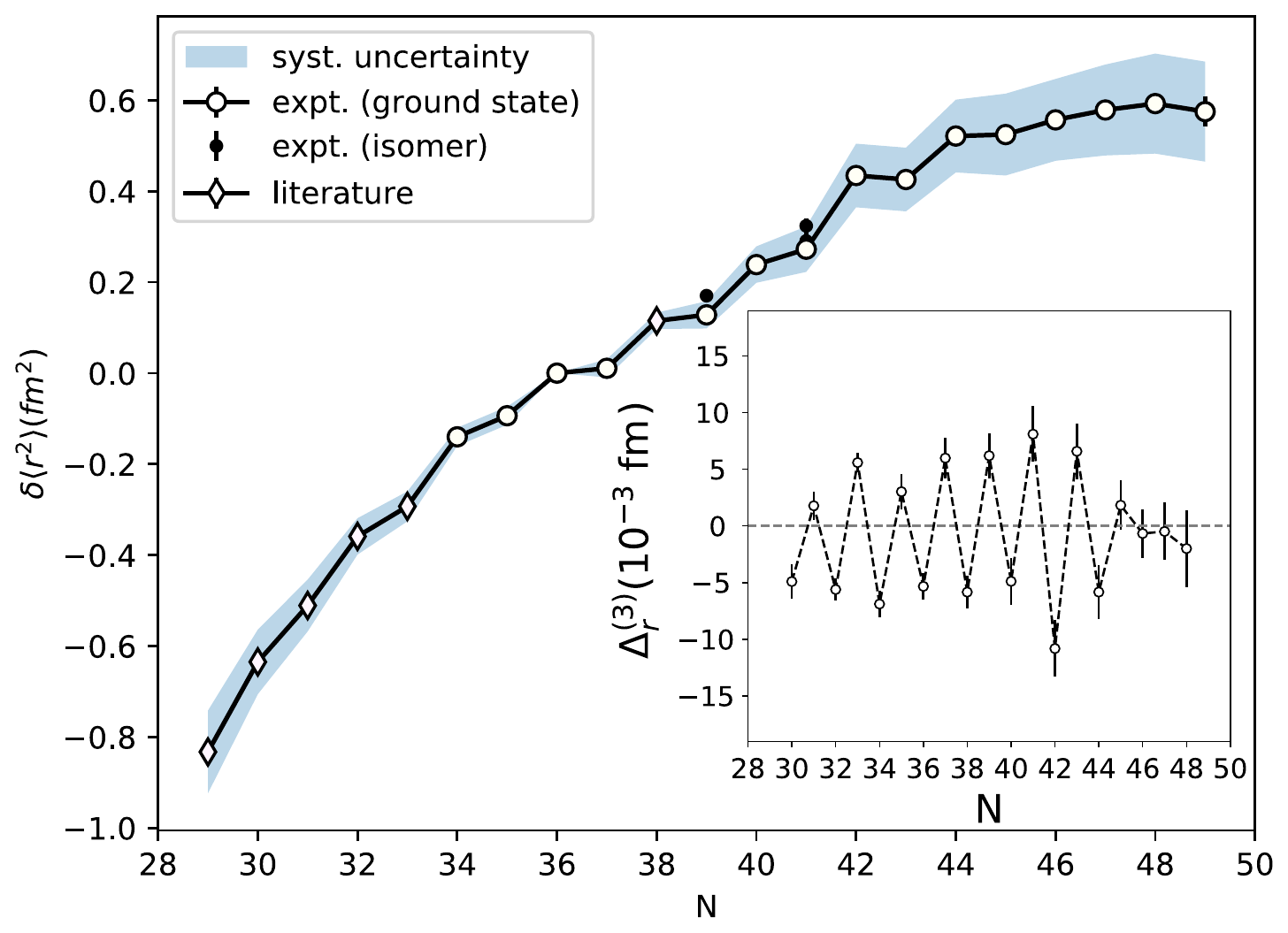}
  \caption{Top: Hyperfine spectrum of $^{78}$Cu. Bottom: Values of $\delta\left\langle r^2 \right\rangle$ for $^{63-66,68-78}$Cu (relative to $^{65}$Cu) obtained in this work alongside the literature values for $^{58-62,67}$Cu. The inset shows the OES of the radii defined by \eqref{OES_radii}.}\label{fig:radii}
\end{figure}

Here, we present data which extends our knowledge of the charge radii of the copper isotopes by another three neutron numbers, up to $^{78}$Cu $(N=49)$. These data represent an important step towards understanding of the nuclear sizes in close proximity to the very neutron-rich doubly magic system of $^{78}$Ni~\cite{Taniuchi2019}. The high efficiency and selectivity of the CRIS technique, developed over the past few years~\cite{deGroote2017}, allows the observation of signals with detection rates of less than 0.05\,ions/s on resonance (see Fig.\,\ref{fig:radii}). Thanks to the ultra-low background rates inherent to the method, these detection rates were sufficient for a successful measurement of the charge radius of $^{78}$Cu in less than one day, while other isotopes typically only require a few hours of beam time. The signal-to-background ratio obtained is similar to those achieved in the state-of-the-art in-source measurements~\cite{marsh2018}, but with much narrow linewidths of $<100$\,MHz, typical for conventional fast-beam collinear laser spectroscopy techniques, thus demonstrating a best-of-both-worlds performance. 

For the purpose of our experiments, radioactive ions were produced at the ISOLDE laboratory at CERN. This was done by impinging 1.4-GeV protons onto a neutron converter, producing  neutrons which in turn induced fission of $^{238}$U atoms within a thick target. Several purification and selection steps were taken to remove unwanted contaminants. First, the copper atoms which diffused out of the target were element-selectively laser-ionised by the ISOLDE RILIS in a hot cavity. The ions were then extracted and accelerated to 30\,keV to prepare them for mass separation. This second selection stage separates ions with different mass-to-charge ratios using a magnetic dipole separator, called the ISOLDE High Resolution Separator. After this, the ions enter into a gas-filled radio-frequency linear Paul trap, ISCOOL, where they accumulated for up to 10\,ms. The ions are then released in a short bunch with a length of $\sim1\,\mu$s, and guided into the CRIS beamline, where they were neutralized through a charge-exchange reaction with a potassium vapor. The non-neutralized fraction of the beam was deflected, such that only the neutralized atoms entered into an ultra-high vacuum region, where they interacted with two pulsed laser beams (for more details, see Methods). The first of these laser systems, tuned to the optical transition at 40114.01\,cm$^{-1}$, resonantly excited the atoms, while the second laser further excited these atoms to an auto-ionising state. Due to the vacuum of $10^{-8}$\,mbar, the non-resonant collisional ionisation rate was less than 1 every 10 minutes for all except the stable $^{63,65}$Cu, creating a quasi background-free measurement. The resulting ions were deflected onto a multi-channel plate detector. As illustrated in the top panel of Fig.~\ref{fig:radii}, by recording the number of ions as a function of the frequency of the first single-mode laser, the hyperfine structure of the copper atoms could be measured.

The changes in the mean-squared charge radii extracted from the hyperfine structure spectra are plotted in the bottom panel of Fig.\,\ref{fig:radii} (white dots), complemented by literature values for $^{58-62,67}$Cu~\cite{Bissell2016} (white diamonds). The radii of the isomeric states are shown with black markers. The shaded area shows the uncertainty due to the atomic parameters (see Methods section for more details). Values of the three-point OES parameter of the radii $\Delta^{(3)}_{r}$, defined as
\begin{equation}
    \Delta^{(3)}_{r} = \frac12 \left(r_{A+1} - 2r_A + r_{A-1}\right),\label{OES_radii}
\end{equation}
are shown in the inset of Fig.\,\ref{fig:radii}. The OES of the radii is quite pronounced near $N=40$, but our new data points reveal a reduction of the OES towards $N=50$, starting at $^{74}$Cu. This is likely to be attributed to the change in the ground-state proton configuration. Indeed, as reflected in the ground-state spins and moments~\cite{Flanagan2009,deGroote2017b}, up to $^{73}$Cu, the odd proton resides dominantly in the $\pi p_{3/2}$ orbital, while from $^{74}$Cu onwards it occupies the $\pi f_{5/2}$ shell. 

We will now demonstrate that modern DFT and the ab-initio VS-IMSRG frameworks can both provide a satisfactory understanding of changes in the charge radii and binding energies of the copper isotopic chain between neutron numbers $N=29$ and $N=49$, down to the scale of the small OES. In the context of the following discussion, it is important to remember that the global (bulk) behavior of nuclear charge radii is governed by the Wigner-Seitz (or box-equivalent) radius $r_0=[3/(4\pi\rho_0)]^{1/3}$, which is given by the nuclear saturation density  $\rho_0$. On the other hand, the local fluctuations in charge radii, including OES, are primarily impacted by the shell structure and many-body correlations. The common interpretation of OES involves various types of polarisations exerted by an odd nucleon, occupying a specific shell-model (or one-quasiparticle) orbital \cite{Xie2019}. In particular, the self-consistent coupling between the neutron pairing field and the proton density provides a coherent understanding of OES of charge radii of spherical nuclei such as semi-magic isotopic chains~\cite{Zawischa1987,FayZ96,fayans2000,Reinhard2017}. 

With measurements now spanning all isotopes between the two exotic doubly-magic systems $^{56,78}$Ni, the copper isotopes represent an ideal laboratory for testing novel theoretical approaches in the medium-mass region. This region of the nuclear chart represents new territory for ab-initio theories based on two- (NN) and three-nucleon (3N) forces derived from chiral effective field theory~\cite{Epel09RMP,Mach11PR}. In general, OES of masses has only been sparsely studied within the context of nuclear forces and many-body methods~\cite{Lesi11gaps,Holt13gaps}. However, the VS-IMSRG~\cite{Tsuk12SM,Stro19ARNPS} has now sufficiently advanced to study most nuclear properties in essentially all open-shell systems below $A=100$, including masses, charge radii, spectroscopy, and electroweak transitions~\cite{Hend18E2}. The presence of a potential sub-shell closure at $N = 40$ and the well-evidenced structural changes due to shell evolution as $N = 50$ is approached~\cite{Flanagan2009} all serve to test such calculations even further. From the side of the DFT calculations, the recently developed Fayans functional, successful in describing the global trends of charge radii in the Sn ($Z=50$) and Ca ($Z=20$) mass regions~\cite{Reinhard2017,hammen2018,gorges2019}, has not been tested in this region of the nuclear chart, nor with data on odd-$Z$ isotopes in general. 

Details on both the DFT and VS-IMSRG calculations can be found in the Methods section, but a few key aspects will be mentioned. The DFT calculations were carried out with the Fayans energy density functional~\cite{Fayans1998}, which -- importantly -- reproduces the microscopic equations of state of symmetric nuclear matter and neutron matter. The inclusion of surface and pairing terms dependent on density gradients has been shown crucial for reproducing (the OES of) the calcium charge radii~\cite{Reinhard2017}. The VS-IMSRG calculations were performed with two sets of NN+3N forces derived from chiral effective field theory, the PWA and 1.8/2.0(EM) interactions of Refs.~\cite{Hebe11fits}. Both are constrained by only 2-, 3-, and 4-body data, with 3N-forces specifically fit to reproduce the $^3$H binding energy and $^4$He charge radius.

\begin{table}[!ht]
\centering
\setlength{\tabcolsep}{3pt}
\caption{Isotope shifts obtained in this work, compared to the available literature~\cite{Bissell2016}. Statistical and atomic-systematic uncertainties are enclosed within the parentheses and square brackets, respectively.}\label{tab:results}
\begin{tabular}{r|ccccc}
    A & I & IS (MHz) &  $\delta\left\langle r^2 \right\rangle^{65, A'}$ (fm$^2$)  & $\delta\left\langle r^2 \right\rangle^{65, A'}_{\text{lit}}$ (fm$^2$) \\[3pt]
    \hline
    63  & 3/2 &  +1055.4(2.6)  & -0.140(7)[20] & -0.148(1)[17]    \\
    64  & 1   &  +508(5)       & -0.09(1)[2]   &  -0.118(3)[13]   \\
    65  & 3/2 &     0          &  0(0)[0]      &   0(0)[0]        \\
    66  & 1   &  -529(5)       & +0.01(1)[2]   &  +0.033(5)[12]   \\
    67  & 3/2 &     -          &    -          &  +0.115(5)[18]   \\
    68  & 1   &  -1498(5)      & +0.13(1)[3]   &  +0.132(5)[31]   \\
    68  & 6   &  -1479(4)      & +0.17(1)[3]   &  +0.191(4)[31]   \\
    69  & 3/2 &  -1937(5)      & +0.24(1)[4]   & +0.237(3)[34]    \\
    70  & 6   &  -2375(7)      & +0.27(2)[5]   &  +0.270(3)[44]   \\
    70  & 3   &  -2390(7)      & +0.29(2)[5]   &  +0.287(11)[44]  \\
    70  & 1   &  -2397(6)      & +0.32(2)[5]   &  +0.323(11)[44]  \\
    71  & 3/2 &  -2787(4)      & +0.44(2)[7]   & +0.407(11)[44]   \\
    72  & 2   &  -3240(7)      & +0.43(2)[7]   &  +0.429(5)[55]   \\
    73  & 3/2 &  -3634(5)      & +0.52(2)[8]   & +0.523(15)[58]   \\
    74  & 2   &  -4057(5)      & +0.53(1)[9]   &  +0.505(18)[72]  \\
    75  & 5/2 &  -4455(4)      & +0.56(1)[9]   & +0.546(21)[80]   \\
    76  & 3   &  -4848(5)      & +0.58(2)[10]  &   -              \\
    77  & 5/2 &  -5234(5)      & +0.59(2)[11]  &   -              \\
    78  & (6) & -5623(10)      & +0.58(3)[11]  &   -              \\
\end{tabular}
\end{table}

\begin{figure}[ht!]
\includegraphics[width=0.9\columnwidth]{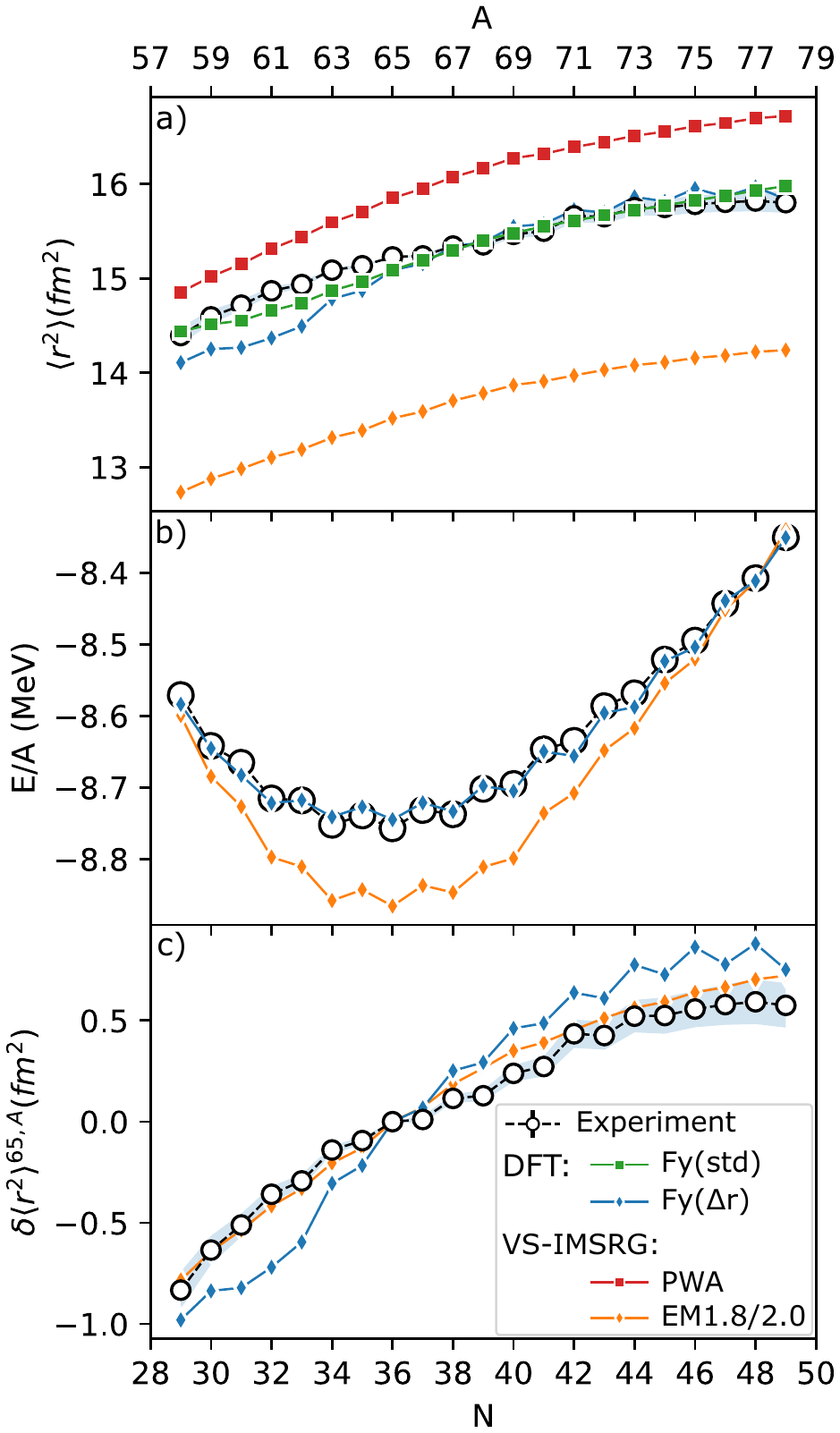}
  \caption{Comparison of experimental squared charge radii (a), binding energies per nucleon (b), and differential squared charge radii (c) with model predictions. For clarity, only Fy($\Delta r$,HFB) and EM1.8/2.0 results  are shown in panels (b) and (c).}\label{fig:abs}
\end{figure}

The absolute charge radii of the copper isotopes are compared to the theoretical calculations in Fig.\,\ref{fig:abs}(a). These total charge radii are obtained using the reference radius~\cite{Angeli2013} $r_{65}= 3.9022(14)$\,fm. Excellent overall agreement was obtained with the two sets of DFT calculations, whereas the ab-initio approaches either generated absolute radii that are too small (EM1.8/2.0) or too large (PWA). This confirms earlier findings~\cite{hagen2016,reinhard2016} that the reproduction of the absolute radii requires a correct prediction of the nuclear saturation density $\rho_0$. While both Fayans functionals used here meet this condition, this is not the case for the interactions used in VS-IMSRG: the EM1.8/2.0 interaction saturates at too high of a density, and the PWA interaction saturates at a somewhat lower density. 

The mismatch between DFT and experiment for the neutron-deficient isotopes is primarily related to the pairing in the $\pi f_{5/2}$ $p$-shell region: as discussed in the Methods, the pairing gradient term which was adjusted using data from the calcium region is too strong in heavier nuclei. We note that a similar reduction around $N=30$ was predicted by the Fayans DF3-a1 DFT calculations for the Ni chain~\cite{Saperstein2011}. For the binding energies per nucleon~\cite{welker2017}, shown in the middle panel of Fig.\,\ref{fig:abs}, the DFT calculations matched particularly well with experiment, with both interactions yielding practically identical results. The VS-IMSRG binding energies appear to overbind in the mid-shell region but nevertheless do  well in terms of the absolute value as well as the general trend.  Figure\,\ref{fig:abs}(c) shows  $\delta\left\langle r^2 \right\rangle^{65, A}$ alongside EM1.8/2.0 VS-IMSRG and the Fy($\Delta r$) calculations. The agreement of the VS-IMSRG calculations with the data is excellent overall except for a small discrepancy near $N=40$.

\begin{figure}[ht!]
  \includegraphics[width=\columnwidth]{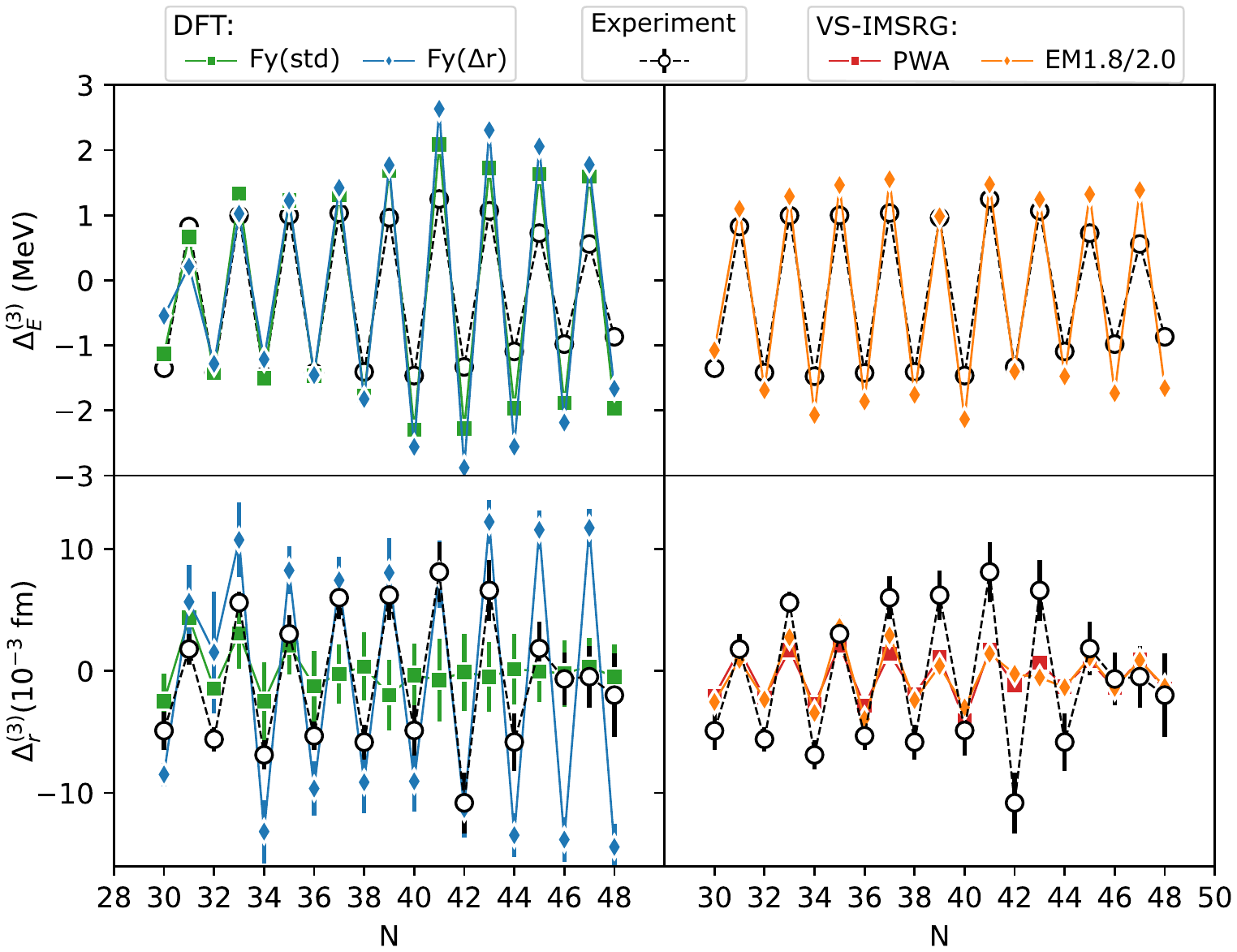}
  \caption{Comparison of the OES of the binding energies (top) and charge radii (bottom) with DFT (left) and ab-initio calculations (right).}\label{fig:stagger}
\end{figure}

Figure\,\ref{fig:stagger} looks at the OES in more detail, by comparing experimental values of $\Delta^{(3)}_{E}$ and $\Delta^{(3)}_{r}$ to theoretical values obtained with DFT  and VS-IMSRG. Note how the reduction in experimental $\Delta^{(3)}_{r}$ near $N=50$ is also visible in $\Delta^{(3)}_{E}$, although less pronounced. This demonstrates that charge radii are more sensitive to the underlying nuclear structure changes in this region, such as to the spin change due to the inversion of the $p_{3/2}$ and $f_{5/2}$ proton orbitals from $N=45$ onwards. The OES in binding energies are reproduced very well with the VS-IMSRG calculations, while it is somewhat overestimated with DFT. For the charge radii, the functional Fy($\Delta r$,HFB) which includes the OES in calcium isotopes as fitting parameters, generates significantly more OES than Fy(std), as expected. Near $N=40$, the agreement for $\Delta^{(3)}_{r}$ is particularly good. The estimated uncertainty of the DFT calculations due to the uniform blocking approximation employed is about $5 \cdot 10^{-3}$\,fm, which covers the small quantitative mismatch of Fy($\Delta r$,HFB) at small $N$, though the deviation at large $N$ remains significant and provides helpful benchmark for further development. 

The VS-IMSRG calculations predict an OES of the right order of magnitude close to the neutron-shell closures $N=28, 50$, but do not reproduce the larger OES in the midshell and close to $N=40$. This is very likely related to the missing proton excitations from the $\pi f_{7/2}$ shell~\cite{deGroote2017b}, which were shown to be important from $N=41$ onwards, but reduced in $^{77,78}$Cu. The observation that the VS-IMSRG performs as well or even better than DFT when it comes to predicting the local features of the radii and energies illustrates that the many-body correlation effects are under better control in VS-IMSRG. This provides an important clue to the microscopic origins of the OES. In particular, the natural emergence of OES of both radii and binding energies from an interaction which is only constrained with 4-body data is encouraging.

In conclusion, we have reported new measurements of the changes in the mean-squared charge radii of very neutron-rich copper isotopes, which were made possible thanks to the high resolution and high sensitivity of the newly developed collinear resonance ionisation method at CERN.  This technique is now available to study exotic isotopes, approaching the nuclear driplines, thus accessing for the first time the anchor points in the nuclear chart that benchmark nuclear theories.  We demonstrated good agreement between our measurements and results from DFT and IMSRG theory. Given the intrinsic complexity of medium-mass systems with odd-$Z$, this represents also a major achievement in nuclear theory, and an important step forward in our global understanding of the nuclear binding energy and charge radius of exotic isotopes. The interplay between the bulk nuclear properties (better captured by DFT) and local variations (better captured by IMSRG) was shown crucial in revealing the microscopic description of the OES effect in radii and binding energies.  The OES emerges naturally from NN+3N interactions derived from chiral effective field theory, constrained to the properties of isotopes with up to four nucleons only, which presents a significant step forward towards a predictive nuclear theory. The comparison with heavier odd-$Z$ systems near the heaviest self-conjugate isotope $^{100}$Sn, which can now be studied experimentally over a range or more than 30 isotopes, are expected to provide the next challenge for nuclear theory. 

\vspace{6pt}

\textbf{Acknowledgements}
We acknowledge the support of the ISOLDE collaboration and technical teams, and the University of Jyv\"{a}skyl\"{a} for the use of the injection-locked cavity. This work was supported by the BriX Research Program No.~P7/12 and FWO-Vlaanderen (Belgium) and GOA 15/010 from KU Leuven, FNPMLS ERC Consolidator Grant no.~648381, the Science and Technology Facilities Council consolidated grant ST/P004423/1 and continuation grant ST/L005794/1,  the EU Seventh Framework through ENSAR2 (654002), and by the Office of Science, U.S. Department of Energy under Award Numbers DE-SC0013365 and DE-SC0018083 (NUCLEI SciDAC-4 collaboration). We acknowledge the financial aid of the Ed Schneiderman Fund at New York University.

\vspace{6pt}

\textbf{Author contributions} R.P.d.G, J.B., C.L.B., M.L.B., T.E.C., T.D.G., G.J.F.S., D.V.F., K.T.F., S.F., R.F.G.R., W.G., \'{A}.K., K.M.L., G.N., S.R., H.H.S., A.R.V., K.D.A.W., S.G.W., Z.Y.X and X.F.Y. performed
the experiment.  R.P.d.G and C.L.B. performed the data analysis and R.P.d.G prepared the
figures. J.D.H. performed the VS-IMSRG calculations. W.N. and P.G.R performed the DFT calculations. R.P.d.G, W.N., P.G.R. and J.H. prepared the manuscript. All authors discussed the results and contributed to the manuscript at all stages.

\newpage

\textbf{METHODS}

\vspace{4pt}

\noindent

\textbf{Laser systems} Copper atoms were laser-ionised using a two-step laser ionisation scheme. Light for the first step was produced using an injection-locked titanium-sapphire laser system jointly developed by the Johannes Gutenberg-Universit\"{a}t Mainz and the University of Jyv\"{a}skyl\"{a}~\cite{Sonnenschein2017}. This laser cavity is built around a titanium-sapphire crystal which is pumped with 532\,nm light produced at a repetition rate of 1\,kHz by a Lee LDP-100MQG pulsed Nd:YAG laser. Through pulsed amplification of cw seed light produced by an M-squared SolsTiS titanium-sapphire laser, narrowband ($\approx$20\,MHz) pulsed laser light was produced at a repetition rate of 1\,kHz. This laser light was then frequency tripled using two nonlinear crystals to produce the required 249\,nm for the resonant excitation step. A maximum of 0.5\,$\mu$J of 249\,nm light was delivered into the CRIS beamline, saturating the resonant step. The wavelength of the scanning laser was recorded by a HighFinesse WSU2 wavemeter every 10\,ms, and used in a feedback loop to stabilize the wavelength to a target value. The wavelength of a temperature-stabilized HeNe laser was simultaneously recorded to evaluate the drift of the wavemeter during the measurements. Resonant ionisation of these excited atoms was achieved using a 314.2444\,nm transition to the $3d^94s(^3D)4d \ ^4P_{3/2}$ auto-ionising state at 71927.28\,cm$^{-1}$, using light produced by a frequency-doubled Spectrolase 4000 pulsed dye laser pumped by a Litron LPY 601 50-100 PIV Nd:YAG laser, at a repetition rate of 100\,Hz. Due to the 344(20)\,ns lifetime of the excited state, the 314\,nm laser pulse could be delayed by 50\,ns, removing lineshape distortions and power broadening effects~\cite{deGroote2017} without appreciable efficiency losses. A small pick-off of the fundamental laser light is sent to a Highfinesse WS6 wavemeter to monitor potential wavelength drifts. The maximum power of this system was 125\,$\mu$J, which was not enough to fully saturate the transition. 

\vspace{6pt}

\textbf{Extraction of charge radii from isotope shifts} The hyperfine parameters and the centroids $\nu_A$ are extracted from the hyperfine spectra by fitting them with correlated Voigt profiles centered at the resonance transition frequencies. This analysis was performed using the SATLAS analysis library~\cite{Gins2018}. More details on the analysis procedure can be found in Ref.~\cite{deGrooteThesis2017}. The isotope shift, $\delta \nu^{65, A'} = \nu_{A} - \nu_{65}$, can be determined from the shift of the centroid of the hyperfine spectrum  of one isotope relative to that of a reference isotope. Such reference measurements of $^{65}$Cu were used to take into account possible changes in the beam energy or  drifts in wavemeter calibration during the four-day experiment. Table \ref{tab:results} displays the isotope shifts obtained in this work. From these isotope shifts, the mean-squared charge radius difference $\delta\left\langle r^2 \right\rangle^{65, A'}$ is obtained from:
\begin{equation}
    \delta \nu^{65, A'} = M\frac{m_{A'} - m_{65}}{m_{65}m_{A'}} + F\delta\left\langle r^2 \right\rangle^{65, A'}, \label{isoshift}
\end{equation}
with $M$ and $F$ being the mass- and field-shift factor, respectively. The atomic factors for the 249\,nm line were determined through a King plot analysis~\cite{deGrooteThesis2017} using the data for the 324.7540\,nm line~\cite{Bissell2016}. Using values of $M_{324}~=~1413(27)$\,GHz\,u and $F_{324}~=~-779(78)$\,MHz\,fm$^{-2}$ obtained from the same reference, we find $M_{249}~=~2284(24)$\,GHz\,u and $F_{249}~=~439(80)$\,MHz\,fm$^{-2}$. The extracted radii are in excellent agreement with those measured in the 324 nm transition (see Table \ref{tab:results}), although a factor of 2-4 less precise due to the lower value of $F_{249}$ as compared to $F_{324}$. 

\vspace{6pt}

\textbf{DFT calculations} In this work, we employ the recently developed parametrisations  Fy(std)~\cite{Reinhard2017} and Fy($\Delta r$,HFB)~\cite{Mil18a,gorges2019}, which differ in the pool of data used for the optimization protocol. Fy(std) uses the standard set of ground state data (binding energies and charge/matter radii of even-even semi-magic isotopes) from Ref.~\cite{kluepfel2009} complemented with OES of energies, defined in an analogous way as Eq.\eqref{OES_radii}. Fy($\Delta r$,HFB) additionally uses the charge radii of the even-odd calcium isotopes. The inclusion of this information significantly increases the strength of the gradient term in the pairing functional (which is absent in other DFT forms and practically inactive in Fy(std)). This was found to be crucial in reproducing the OES of charge radii in the Ca isotopes~\cite{Reinhard2017}. However, this parameterization overestimates the OES in the heavier masses~\cite{hammen2018} and the kink in charge radii around $^{132}$Sn~\cite{gorges2019}. For all practical details pertaining to our Fayans-DFT calculations, we refer the reader to Ref.~\cite{Mil18a}.

\vspace{6pt}

\textbf{IM-SRG calculations} Working in an initial harmonic-oscillator basis of 13 major shells, we first transform to the Hartree-Fock basis, then use the Magnus prescription~\cite{Morr15Magnus} of the IMSRG to generate approximate unitary transformations to decouple first the given core energy, then as well as a specific valence-space Hamiltonian from the full $A$-body Hamiltonian. In the IMSRG(2) approximation used in this work here, all operators are truncated at the two-body level. In addition we capture effects of 3N forces between valence nucleons via the ensemble normal ordering procedure of Ref.~\cite{Stro17ENO}. The same transformation is then applied to the intrinsic proton mean-squared radius operator, with appropriate corrections to obtain core and valence-space (hence absolute) charge radii. We take $^{56}$Ni as our core reference state, with a valence space comprised of the proton and neutron $p_{3/2}$, $p_{1/2}$, $f_{5/2}$, and $g_{9/2}$ single-particle orbits. Finally, using the NUSHELLX@MSU shell model code~\cite{Brow14NuShellX}, we diagonalize the valence-space Hamiltonian to obtain ground- (and excited-) state energies, as well as expectation values for the charge radius operator, where induced two-body corrections are included naturally in the VS-IMSRG approach.

\bibliography{biblio}

\begin{thebibliography}{10}
\expandafter\ifx\csname url\endcsname\relax
  \def\url#1{\texttt{#1}}\fi
\expandafter\ifx\csname urlprefix\endcsname\relax\def\urlprefix{URL }\fi
\providecommand{\bibinfo}[2]{#2}
\providecommand{\eprint}[2][]{\url{#2}}

\bibitem{angeli2013}
\bibinfo{author}{Angeli, I.} \& \bibinfo{author}{Marinova, K.}
\newblock \bibinfo{title}{Table of experimental nuclear ground state charge
  radii: An update}.
\newblock \emph{\bibinfo{journal}{At. Data Nucl. Data Tables}}
  \textbf{\bibinfo{volume}{99}}, \bibinfo{pages}{69--95}
  (\bibinfo{year}{2013}).
\newblock \urlprefix\url{https://doi.org/10.1016/j.adt.2011.12.006}.

\bibitem{reinhard2016}
\bibinfo{author}{Reinhard, P.-G.} \& \bibinfo{author}{Nazarewicz, W.}
\newblock \bibinfo{title}{Nuclear charge and neutron radii and nuclear matter:
  Trend analysis in {Skyrme} density-functional-theory approach}.
\newblock \emph{\bibinfo{journal}{Phys. Rev. C}} \textbf{\bibinfo{volume}{93}},
  \bibinfo{pages}{051303} (\bibinfo{year}{2016}).
\newblock \urlprefix\url{https://doi.org/10.1103/PhysRevC.93.051303}.

\bibitem{hammen2018}
\bibinfo{author}{Hammen, M.} \emph{et~al.}
\newblock \bibinfo{title}{From calcium to cadmium: Testing the pairing
  functional through charge radii measurements of {$^{100-130}$Cd}}.
\newblock \emph{\bibinfo{journal}{Phys. Rev. Lett.}}
  \textbf{\bibinfo{volume}{121}}, \bibinfo{pages}{102501}
  (\bibinfo{year}{2018}).
\newblock \urlprefix\url{https://doi.org/10.1103/PhysRevLett.121.102501}.

\bibitem{gorges2019}
\bibinfo{author}{Gorges, C.} \emph{et~al.}
\newblock \bibinfo{title}{Laser spectroscopy of neutron-rich tin isotopes: A
  discontinuity in charge radii across the ${N= 82}$ shell closure}.
\newblock \emph{\bibinfo{journal}{Phys. Rev. Lett.}}
  \textbf{\bibinfo{volume}{122}}, \bibinfo{pages}{192502}
  (\bibinfo{year}{2019}).
\newblock \urlprefix\url{https://doi.org/10.1103/PhysRevLett.122.192502}.

\bibitem{Reinhard2017}
\bibinfo{author}{Reinhard, P.-G.} \& \bibinfo{author}{Nazarewicz, W.}
\newblock \bibinfo{title}{Toward a global description of nuclear charge radii:
  Exploring the fayans energy density functional}.
\newblock \emph{\bibinfo{journal}{Phys. Rev. C}} \textbf{\bibinfo{volume}{95}},
  \bibinfo{pages}{064328} (\bibinfo{year}{2017}).
\newblock \urlprefix\url{https://doi.org/10.1103/PhysRevC.95.064328}.

\bibitem{Mil18a}
\bibinfo{author}{Miller, A.~J.} \emph{et~al.}
\newblock \bibinfo{title}{Proton superfluidity and charge radii in proton-rich
  calcium isotopes}.
\newblock \emph{\bibinfo{journal}{Nature Phys.}} \textbf{\bibinfo{volume}{15}},
  \bibinfo{pages}{1745--2473} (\bibinfo{year}{2019}).
\newblock \urlprefix\url{https://doi.org/10.1038/s41567-019-0416-9}.

\bibitem{Tsuk12SM}
\bibinfo{author}{Tsukiyama, K.}, \bibinfo{author}{Bogner, S.} \&
  \bibinfo{author}{Schwenk, A.}
\newblock \bibinfo{title}{In-medium similarity renormalization group for
  open-shell nuclei}.
\newblock \emph{\bibinfo{journal}{Phys. Rev. C}} \textbf{\bibinfo{volume}{85}},
  \bibinfo{pages}{061304(R)} (\bibinfo{year}{2012}).
\newblock \urlprefix\url{https://doi.org/10.1103/PhysRevC.85.061304}.

\bibitem{Stro19ARNPS}
\bibinfo{author}{Stroberg, S.~R.}, \bibinfo{author}{Bogner, S.~K.},
  \bibinfo{author}{Hergert, H.} \& \bibinfo{author}{Holt, J.~D.}
\newblock \bibinfo{title}{Nonempirical interactions for the nuclear shell
  model: An update}.
\newblock \emph{\bibinfo{journal}{Annual Review of Nuclear and Particle
  Science}} \textbf{\bibinfo{volume}{69}}, \bibinfo{pages}{307--362}
  (\bibinfo{year}{2019}).
\newblock \urlprefix\url{https://doi.org/10.1146/annurev-nucl-101917-021120}.

\bibitem{bissell2016}
\bibinfo{author}{Bissell, M.~L.} \emph{et~al.}
\newblock \bibinfo{title}{Cu charge radii reveal a weak sub-shell effect at
  {$N= 40$}}.
\newblock \emph{\bibinfo{journal}{Phys. Rev. C}} \textbf{\bibinfo{volume}{93}},
  \bibinfo{pages}{064318} (\bibinfo{year}{2016}).
\newblock \urlprefix\url{https://doi.org/10.1103/PhysRevC.93.064318}.

\bibitem{deGroote2017}
\bibinfo{author}{de~Groote, R.~P.} \emph{et~al.}
\newblock \bibinfo{title}{Efficient, high-resolution resonance laser ionization
  spectroscopy using weak transitions to long-lived excited states}.
\newblock \emph{\bibinfo{journal}{Phys. Rev. A}} \textbf{\bibinfo{volume}{95}},
  \bibinfo{pages}{032502} (\bibinfo{year}{2017}).
\newblock \urlprefix\url{https://link.aps.org/doi/10.1103/PhysRevA.95.032502}.

\bibitem{Taniuchi2019}
\bibinfo{author}{Taniuchi, R.} \emph{et~al.}
\newblock \bibinfo{title}{{$^{78}$Ni} revealed as a doubly magic stronghold
  against nuclear deformation}.
\newblock \emph{\bibinfo{journal}{Nature}} \textbf{\bibinfo{volume}{569}},
  \bibinfo{pages}{53--58} (\bibinfo{year}{2019}).
\newblock \urlprefix\url{https://doi.org/10.1038/s41586-019-1155-x}.

\bibitem{marsh2018}
\bibinfo{author}{Marsh, B.~A.} \emph{et~al.}
\newblock \bibinfo{title}{Characterization of the shape-staggering effect in
  mercury nuclei}.
\newblock \emph{\bibinfo{journal}{Nature Phys.}} \textbf{\bibinfo{volume}{14}},
  \bibinfo{pages}{1163--1167} (\bibinfo{year}{2018}).
\newblock \urlprefix\url{https://doi.org/10.1038/s41567-018-0292-8}.

\bibitem{Flanagan2009}
\bibinfo{author}{Flanagan, K.~T.} \emph{et~al.}
\newblock \bibinfo{title}{Nuclear spins and magnetic moments of
  {$^{71,73,75}$Cu}: Inversion of {$\pi2p_{3/2}$} and {$\pi1f_{5/2}$} levels in
  {$^{75}$Cu}}.
\newblock \emph{\bibinfo{journal}{Phys. Rev. Lett.}}
  \textbf{\bibinfo{volume}{103}}, \bibinfo{pages}{142501}
  (\bibinfo{year}{2009}).
\newblock \urlprefix\url{https://doi.org/10.1103/PhysRevLett.103.142501}.

\bibitem{deGroote2017b}
\bibinfo{author}{de~Groote, R.~P.} \emph{et~al.}
\newblock \bibinfo{title}{Dipole and quadrupole moments of {$^{73-78}$Cu} as a
  test of the robustness of the {$Z= 28$} shell closure near {78Ni}}.
\newblock \emph{\bibinfo{journal}{Phys. Rev. C}} \textbf{\bibinfo{volume}{96}},
  \bibinfo{pages}{041302} (\bibinfo{year}{2017}).
\newblock \urlprefix\url{https://doi.org/10.1103/PhysRevC.96.041302}.

\bibitem{Xie2019}
\bibinfo{author}{Xie, L.} \emph{et~al.}
\newblock \bibinfo{title}{Nuclear charge radii of {$^{62-80}$Zn} and their
  dependence on cross-shell proton excitations}.
\newblock \emph{\bibinfo{journal}{Physics Letters B}}
  \textbf{\bibinfo{volume}{797}}, \bibinfo{pages}{134805}
  (\bibinfo{year}{2019}).
\newblock \urlprefix\url{https://doi.org/10.1016/j.physletb.2019.134805}.

\bibitem{Zawischa1987}
\bibinfo{author}{Zawischa, D.}, \bibinfo{author}{Regge, U.} \&
  \bibinfo{author}{Stapel, R.}
\newblock \bibinfo{title}{Effective interaction and the staggering of nuclear
  charge radii}.
\newblock \emph{\bibinfo{journal}{Phys. Lett. B}}
  \textbf{\bibinfo{volume}{185}}, \bibinfo{pages}{299--303}
  (\bibinfo{year}{1987}).
\newblock \urlprefix\url{https://doi.org/10.1016/0370-2693(87)91003-3}.

\bibitem{FayZ96}
\bibinfo{author}{Fayans, S.} \& \bibinfo{author}{Zawischa, D.}
\newblock \bibinfo{title}{Towards a better parametrization of the nuclear
  pairing force: density dependence with gradient term}.
\newblock \emph{\bibinfo{journal}{Phys. Lett. B}}
  \textbf{\bibinfo{volume}{383}}, \bibinfo{pages}{19--23}
  (\bibinfo{year}{1996}).
\newblock
  \urlprefix\url{http://www.sciencedirect.com/science/article/pii/0370269396007162}.

\bibitem{fayans2000}
\bibinfo{author}{Fayans, S.}, \bibinfo{author}{Tolokonnikov, S.},
  \bibinfo{author}{Trykov, E.} \& \bibinfo{author}{Zawischa, D.}
\newblock \bibinfo{title}{Nuclear isotope shifts within the local
  energy-density functional approach}.
\newblock \emph{\bibinfo{journal}{Nucl. Phys. A}}
  \textbf{\bibinfo{volume}{676}}, \bibinfo{pages}{49--119}
  (\bibinfo{year}{2000}).
\newblock \urlprefix\url{https://doi.org/10.1016/S0375-9474(00)00192-5}.

\bibitem{Epel09RMP}
\bibinfo{author}{Epelbaum, E.}, \bibinfo{author}{Hammer, H.-W.} \&
  \bibinfo{author}{Mei{\ss}ner, U.-G.}
\newblock \bibinfo{title}{{Modern Theory of Nuclear Forces}}.
\newblock \emph{\bibinfo{journal}{Rev. Mod. Phys.}}
  \textbf{\bibinfo{volume}{81}}, \bibinfo{pages}{1773--1825}
  (\bibinfo{year}{2009}).
\newblock \urlprefix\url{https://doi.org/10.1103/RevModPhys.81.1773}.

\bibitem{Mach11PR}
\bibinfo{author}{Machleidt, R.} \& \bibinfo{author}{Entem, D.~R.}
\newblock \bibinfo{title}{{Chiral effective field theory and nuclear forces}}.
\newblock \emph{\bibinfo{journal}{Phys. Rep.}} \textbf{\bibinfo{volume}{503}},
  \bibinfo{pages}{1--75} (\bibinfo{year}{2011}).
\newblock \urlprefix\url{https://doi.org/10.1016/j.physrep.2011.02.001}.

\bibitem{Lesi11gaps}
\bibinfo{author}{Lesinski, T.}, \bibinfo{author}{Hebeler, K.},
  \bibinfo{author}{Duguet, T.} \& \bibinfo{author}{Schwenk, A.}
\newblock \bibinfo{title}{{Chiral three-nucleon forces and pairing in nuclei}}.
\newblock \emph{\bibinfo{journal}{J. Phys. G}} \textbf{\bibinfo{volume}{39}},
  \bibinfo{pages}{015108} (\bibinfo{year}{2012}).
\newblock \urlprefix\url{https://doi.org/10.1088/0954-3899/39/1/015108}.

\bibitem{Holt13gaps}
\bibinfo{author}{Holt, J.~D.}, \bibinfo{author}{Menendez, J.} \&
  \bibinfo{author}{Schwenk, A.}
\newblock \bibinfo{title}{{The role of three-nucleon forces and many-body
  processes in nuclear pairing}}.
\newblock \emph{\bibinfo{journal}{J. Phys. G}} \textbf{\bibinfo{volume}{40}},
  \bibinfo{pages}{075105} (\bibinfo{year}{2013}).
\newblock \urlprefix\url{https://doi.org/10.1088/0954-3899/40/7/075105}.

\bibitem{Hend18E2}
\bibinfo{author}{Henderson, J.} \emph{et~al.}
\newblock \bibinfo{title}{{Testing microscopically derived descriptions of
  nuclear collectivity: Coulomb excitation of $^{22}$Mg}}.
\newblock \emph{\bibinfo{journal}{Phys. Lett. B}}
  \textbf{\bibinfo{volume}{782}}, \bibinfo{pages}{468--473}
  (\bibinfo{year}{2018}).
\newblock \urlprefix\url{https://doi.org/10.1016/j.physletb.2018.05.064}.

\bibitem{Fayans1998}
\bibinfo{author}{Fayans, S.~A.}
\newblock \bibinfo{title}{Towards a universal nuclear density functional}.
\newblock \emph{\bibinfo{journal}{JETP Lett.}} \textbf{\bibinfo{volume}{68}},
  \bibinfo{pages}{169--174} (\bibinfo{year}{1998}).
\newblock \urlprefix\url{http://dx.doi.org/10.1134/1.567841}.

\bibitem{Hebe11fits}
\bibinfo{author}{Hebeler, K.}, \bibinfo{author}{Bogner, S.~K.},
  \bibinfo{author}{Furnstahl, R.~J.}, \bibinfo{author}{Nogga, A.} \&
  \bibinfo{author}{Schwenk, A.}
\newblock \bibinfo{title}{{Improved nuclear matter calculations from chiral
  low-momentum interactions}}.
\newblock \emph{\bibinfo{journal}{Phys. Rev. C}} \textbf{\bibinfo{volume}{83}},
  \bibinfo{pages}{031301(R)} (\bibinfo{year}{2011}).
\newblock \urlprefix\url{https://doi.org/10.1103/PhysRevC.83.031301}.

\bibitem{hagen2016}
\bibinfo{author}{Hagen, G.} \emph{et~al.}
\newblock \bibinfo{title}{Neutron and weak-charge distributions of the
  {$^{48}$Ca} nucleus}.
\newblock \emph{\bibinfo{journal}{Nature Phys.}} \textbf{\bibinfo{volume}{12}},
  \bibinfo{pages}{186--190} (\bibinfo{year}{2016}).
\newblock \urlprefix\url{https://doi.org/10.1038/nphys3529}.

\bibitem{ame2017}
\bibinfo{author}{Wang, M.} \emph{et~al.}
\newblock \bibinfo{title}{The {AME}2016 atomic mass evaluation ({II}). tables,
  graphs and references}.
\newblock \emph{\bibinfo{journal}{Chin. Phys. C}}
  \textbf{\bibinfo{volume}{41}}, \bibinfo{pages}{030003}
  (\bibinfo{year}{2017}).
\newblock
  \urlprefix\url{https://doi.org/10.1088%2F1674-1137%2F41%2F3%2F030003}.

\bibitem{welker2017}
\bibinfo{author}{Welker, A.} \emph{et~al.}
\newblock \bibinfo{title}{Binding energy of {$^{79}$Cu}: Probing the structure
  of the doubly magic {78Ni} from only one proton away}.
\newblock \emph{\bibinfo{journal}{Phys. Rev. Lett.}}
  \textbf{\bibinfo{volume}{119}}, \bibinfo{pages}{192502}
  (\bibinfo{year}{2017}).
\newblock \urlprefix\url{https://doi.org/10.1103/PhysRevLett.119.192502}.

\bibitem{Saperstein2011}
\bibinfo{author}{Saperstein, E.~E.} \& \bibinfo{author}{Tolokonnikov, S.~V.}
\newblock \bibinfo{title}{Self-consistent theory of finite fermi systems and
  radii of nuclei}.
\newblock \emph{\bibinfo{journal}{Physics of Atomic Nuclei}}
  \textbf{\bibinfo{volume}{74}}, \bibinfo{pages}{1277--1297}
  (\bibinfo{year}{2011}).
\newblock \urlprefix\url{https://doi.org/10.1134/S1063778811090109}.

\bibitem{Sonnenschein2017}
\bibinfo{author}{Sonnenschein, V.} \emph{et~al.}
\newblock \bibinfo{title}{Characterization of a pulsed injection-locked
  {Ti:sapphire} laser and its application to high resolution resonance
  ionization spectroscopy of copper}.
\newblock \emph{\bibinfo{journal}{Laser Physics}}
  \textbf{\bibinfo{volume}{27}}, \bibinfo{pages}{085701}
  (\bibinfo{year}{2017}).
\newblock \urlprefix\url{https://doi.org/10.1088/1555-6611/aa7834}.

\bibitem{Gins2018}
\bibinfo{author}{Gins, W.} \emph{et~al.}
\newblock \bibinfo{title}{Analysis of counting data: Development of the satlas
  python package}.
\newblock \emph{\bibinfo{journal}{Comput. Phys. Comm.}}
  \textbf{\bibinfo{volume}{222}}, \bibinfo{pages}{286--294}
  (\bibinfo{year}{2018}).
\newblock \urlprefix\url{https://doi.org/10.1016/j.cpc.2017.09.012}.

\bibitem{deGrooteThesis2017}
\bibinfo{author}{de~Groote, R.~P.}
\newblock \emph{\bibinfo{title}{{High resolution collinear resonance ionization
  spectroscopy of neutron-rich $^{76,77,78}$Cu isotopes}}}.
\newblock Ph.D. thesis, \bibinfo{school}{KU Leuven} (\bibinfo{year}{2017}).
\newblock \urlprefix\url{https://cds.cern.ch/record/2285661}.

\bibitem{kluepfel2009}
\bibinfo{author}{Kl{\"u}pfel, P.}, \bibinfo{author}{Reinhard, P.-G.},
  \bibinfo{author}{B{\"u}rvenich, T.} \& \bibinfo{author}{Maruhn, J.}
\newblock \bibinfo{title}{Variations on a theme by {Skyrme}: A systematic study
  of adjustments of model parameters}.
\newblock \emph{\bibinfo{journal}{Phys. Rev. C}} \textbf{\bibinfo{volume}{79}},
  \bibinfo{pages}{034310} (\bibinfo{year}{2009}).
\newblock \urlprefix\url{https://doi.org/10.1103/PhysRevC.79.034310}.

\bibitem{Morr15Magnus}
\bibinfo{author}{Morris, T.~D.}, \bibinfo{author}{Parzuchowski, N.~M.} \&
  \bibinfo{author}{Bogner, S.~K.}
\newblock \bibinfo{title}{Magnus expansion and in-medium similarity
  renormalization group}.
\newblock \emph{\bibinfo{journal}{Phys. Rev. C}} \textbf{\bibinfo{volume}{92}},
  \bibinfo{pages}{034331} (\bibinfo{year}{2015}).
\newblock \urlprefix\url{https://link.aps.org/doi/10.1103/PhysRevC.92.034331}.

\bibitem{Stro17ENO}
\bibinfo{author}{Stroberg, S.~R.} \emph{et~al.}
\newblock \bibinfo{title}{{Nucleus-dependent valence-space approach to nuclear
  structure}}.
\newblock \emph{\bibinfo{journal}{Phys. Rev. Lett.}}
  \textbf{\bibinfo{volume}{118}}, \bibinfo{pages}{032502}
  (\bibinfo{year}{2017}).
\newblock \urlprefix\url{https://doi.org/10.1103/PhysRevLett.118.032502}.

\bibitem{Brow14NuShellX}
\bibinfo{author}{Brown, B.~A.} \& \bibinfo{author}{Rae, W. D.~M.}
\newblock \bibinfo{title}{{The Shell-Model Code NuShellX@MSU}}.
\newblock \emph{\bibinfo{journal}{Nuclear Data Sheets}}
  \textbf{\bibinfo{volume}{120}}, \bibinfo{pages}{115--118}
  (\bibinfo{year}{2014}).
\newblock \urlprefix\url{https://doi.org/10.1016/j.nds.2014.07.022}.

\end{thebibliography}

\cleardoublepage

\end{document}